# On Certain Analytical Representations of Cellular Automata

Theophanes E. Raptis[abc]

[a]National Center for Science and Research "Demokritos", Division of Applied Technologies, Computational Applications Group, Athens, Greece.

[b]University of Athens, Department of Chemistry, Laboratory of Physical Chemistry, Athens, Greece

[c]University of Peloponnese, Informatics and Telecommunications Dept., Tripolis, Greece

**Abstract:** We extend a previously introduced semi-analytical representation of a decomposition of CA dynamics in arbitrary dimensions and neighborhood schemes via the use of certain universal maps in which CA rule vectors are derivable from the equivalent of superpotentials. The results justify the search for alternative analog models of computation and their possible physical connections.

Keywords: Automata, Fourier analysis, Computational Structures

**Introduction**

The last decades, the field of Cellular Automata (CA) had a rapid expansion after its initial revival with the seminal Wolfram's paper [1] both due to the increase of computational power allowing exploration of large configurations in sufficient time depth as well as the meeting with other fields like Machine Learning (ML) and Artificial Intelligence (AI) [2], [3], [4] where it delivered promising results. Since many models in the field appear to be capable of universal computation, the author attempted a reformulation in a recent work [5] which would allow for ideally dissipationless computation by special encoding methods that would signal a return to hybrid analog models of computation in a spectral domain.

The methods presented there are valid for arbitrary dimensions via a one dimensional reduction technique as well as time-varying neighborhood topologies which allow the generic decomposition of any such CA model into a cascade of a linear and a non-linear filter (LN). The linear filter is then realized as a circulant convolution filter which is naturally diagonalized via DFT matrices. Leaving aside the details of encodings associated with possible dissipationless realizations, it is equally interesting to examine the possibility of further reduction of the non-linear part in its associated spectral domain. This becomes possible with a new method of reducing the Fourier transform of composite functions that appeared in [6].

To enable direct application of the technique it is imperative to find analytic expressions for the non-linear part of the original decomposition method in [5]. In a previous work [7], Garcia-Morales presented a kind of arithmetized universal map based on a polynomial representation of the rule space which was also generic and used it to extract several interesting properties of CA [8], [9]. In what follows we will present alternative fully analytical representations that are better adapted in the compact LN decomposition. In the next section three equivalent functional schemes for universal maps are explored allowing the translation of the original dynamics into analytical forms, at least one of which presents an exceptional characteristic for actual experimental realizations via holographic techniques. In section 3, a full exploration of the spectral decomposition of the functional composites constructed in the previous section is attempted. Last section concludes discussing various options for further research in the area.

**Alternative analytical representations**

Repeating standard definitions, we restrict attention to CAs defined on regular $D$-dimensional Euclidean lattices via a tuple $\{v_R \Sigma,, N, L^D\}$ comprising a $D$-dimensional lattice, a neighborhood definition $N$, a tape alphabet $\Sigma$ vector in some alphabet basis $b$ ("*radix*") and a rule integer via the standard Wolfram encoding giving rise to an associated rule as $\mathbf{r} = (\sigma_0, \sigma_1, ..., \sigma_{\#(N)})$ where $\#(N) = b^{|N|}$ the "*capacity*" or the number of all possible states of a given neighborhood topology lexicographically ordered when any such comprises |$N$| cells.

The basic reduction method consists in reading remote neighbors in the same way arrays are indexed in RAM via a zig-zag motion. This results in all elements being reduced to a single one dimensional array and referencing is to be done with a special matrix which shall be called heretofore, the "*Interaction Kernel*". The function of the matrix vector product is similar to the original *B-Calculus* of [7] that is, to extract the addressing information or digit pointer to one of the rule vector positions corresponding to any and all possible neighborhood configurations. Since this can only be done in a 1-1 fashion via the polynomial representation the kernel matrix elements are always composed of successive powers of the associated radix. Due to translational invariance of this scheme, the resulting matrix is always circulant or at most piecewise circulant depending on the choice of boundary conditions. In the simplest of cases, symmetric neighborhoods defined through a radius entail a Kronecker factorization of the kernel into sub-matrices acting separately on each dimension while the global diagonalizing matrices are the Kronecker products of the associated DFT Vandermond matrices for the relevant subspaces.

Introducing an intermediate *"activation field"* $h_t$ of total length $L^D$ at any discrete time-step with a $L^D$ x $L^D$ kernel as well as the tape state vector $S_t$, the LN decomposition becomes

$$\vec{h}_{t+1} = \hat{K} \cdot \vec{S}_t$$
$$\vec{S}_{t+1} = R(\mathbf{r}, \vec{h}_{t+1}) \tag{1}$$

The second map $R$ in (1) is used to denote the non-linear part of referencing of the original rule vector. There is a variety of methods that this could be accomplished, the simplest being to perform a semi-linearization by introducing an auxiliary $L^D \times b^{|N|}$ binary indicator matrix as

$$\vec{S}_{t+1} = \hat{R}\left(2^{\vec{h}_{t+1}}\right) \cdot \vec{r} = \hat{R}\left(2^{\hat{K} \cdot \vec{S}_t}\right) \cdot \vec{r} \tag{2}$$

Each row of $R$ is then a binary shift of order defined by the activation field from the binary decoding of $2^{h_t^i}$. One meets again the exponential nature of the chain of maps which was already mentioned in [5] as analogous to a Djikstra hyper-operation or power tower of maps. While clear in its meaning, this last expression does not offers any possibility of diagonalizing both stages.

Instead we may take advantage of some functional expressions based on congruences that allow arbitrary decoders. The basis function for any such can be given with the aid of an asymmetric 2-periodic square pulse function. Using $\Theta(x)$ for the Heaviside function we define

$$\eta(v; p_0, p_1) = \Theta[\mathrm{mod}(v, p_0 + p_1) - p_0] \qquad (3)$$

In (3), the two repeating sub-periods for 0 and 1 intervals form the total period $p = p_0 + p_1$. The advantage of (3) is stands for a direct analog circuit implementation as a discretely sampled voltage controlled sawtooth generator followed by a variable threshold comparator. It is now possible to derive a division-free binary decoder with a reduced parameter set $p_0 = p_1 = 2^l$ where $l$ stands for each digit's siginificance level due to the equivalence of lexicographically ordered sets with periodic counters of an exponential sequence of periods. Hence, we may rewrite (2) in the form

$$\vec{S}_{t+1} = \eta\left(\vec{v}_R, \vec{h}_{t+1}\right) = \eta\left(\vec{v}_R, \hat{K} \cdot \vec{S}_t\right) \qquad (4)$$

In (4), we have the equivalence $\eta(v, h) \cong \eta(v; 2^h, 2^h)$. We also avoid the use of any rule vector using the original Wolfram encoding directly. We also assumed a vectorized implementation with the $\eta$ as a componentwise operator acting on each and every element of the cell state array. While (4) is valid for binary automata it is relatively easy to generalize the "*eta*" function for arbitrary alphabets using a superposition property as

$$\eta_b(v, \mu) = \sum_{k=0}^{b-1} \eta\left(v, kb^{\mu-1}, (b-k)b^{\mu-1}\right) \qquad (5)$$

In (5), there is an accumulator of ones for the reconstruction of each symbol in any lexicographically ordered string set inside $[0,\ldots, b^L]$ while, all periods form a restricted *2*-integer partition of the scaling factor $b^{\mu-1}$.

Since the use of congruencies requires the presence of at least one number theoretic function we have not yet reached a purely analytical representation. To proceed further it is required to review the role of the rule vector or of its integer encoding by making it the derivative of another quantity thus introducing an analogy with what is often called a "*superpotential*" in physical models. There are at least two ways to do this using cumulant forms that are discrete analogs of integral forms. Each has a separate interest regarding possible actual implementations and applications.

We first present the simplest possible form that allows rewriting (2) in a discrete derivative form. Let then *c* be a cumulant vector with $b^{|N|}+1$ elements

$$c_{i+1} = \sum_{i=0}^{\#(N)} r_i, \quad c_0 = 0 \tag{6}$$

Evidently, the original dynamics can now take the (componentwise) form

$$\vec{S}_{t+1} = c(\vec{h}_{t+1}) - c(\vec{h}_{t+1} - 1) \tag{7}$$

Since the superpotential rule function is of a staircase nature, one can always turn it into a smooth form via some spline or other appropriate interpolator.

The second cumulant form is slightly more complicated in an effort to imitate the periodicities of congruencies and is based on the observation that at least a binary rule can easily be reproduced from a different integral form *c* as $r_i = \mathrm{mod}(c_i, 2)$ hence we obtain the superpotential elements as

$$c_i = 2n + r_i, \quad n = 0,1,...2^{|N|} - 1 \tag{8}$$

The same rule can easily be generalized for higher alphabets. In figure 1, both cases are shown as contour maps over the lexicographically ordered set of all possible elementary binary rule strings. It should be stressed that all such sets appear self-similar due to the tree-like structure inherited from the multi-periodic nature of the counter set of constructors for such string sets.

Keeping the binary case as an example, it is possible to mimic the effect of a *modulo* congruency with a continuous harmonic function. Assuming a continuous embedding of the cell lattice with a spacing *δx*, one can introduce a continuous $x = n\delta x$ variable and modify the superpotential form as $\mathbf{c} \to \mathbf{c}' = (\mathbf{c} - 2\mathbf{r} + 1)3\pi/2$ to alternate odd and even positions and obtain the final result as a phase modulation scheme of the form

$$\vec{S}_{t+1} = \Theta\left(\sin\left(k_0 x + c(\vec{h}_{t+1})\right)\right) \tag{9}$$

In actual applications one may approximate the Heavyside function with a sufficiently steep sigmoid approximation like a hyperbolic tangent. In any such expression it is always assumed that $x \neq 0$ to avoid problems with principal values in spectral representations. Notably, (9) is further decomposable as

$$\vec{S}_{t+1} = \Theta\left(\sin(k_0 x)\cos\left(c(\vec{h}_{t+1})\right) + \cos(k_0 x)\sin\left(c(\vec{h}_{t+1})\right)\right) \quad (10)$$

In this latter form one can recognize the superposition of two phase shifted pictures of the same interference pattern with each term comprising the equivalent of a reference beam and a modulated one from an appropriate grating realizing the $c(\vec{h}_{t+1})$ term.

**Spectral decompositions**

From the general properties of circulant convolution filters and denoting with a tilde the DFT of each array we also have the alternative form of the first of (1) as

$$\tilde{\vec{h}}_{t+1} = \hat{Z}_K \cdot \tilde{\vec{S}}_t \quad (11)$$

The eigenvalues in are directly given from the DFT of the first defining row for any circulant matrix. This operation is also known in experimental optics as the "*4-f setup*" [10] since it can be realized with two lenses of focal depths *f* as an example of natural computing via diffractive optics. The second stage becomes then

$$\tilde{\vec{S}}_{t+1} = (F \circ R)(\vec{h}_{t+1}) \quad (12)$$

The transformation matrix *F* is generic and it may deviate from a simple Vandermond one for arbitrary choices of boundary conditions in the multi-dimensional case. For this reason we shall heretofore restrict attention to 1*D* models. Since some of them have been already proven capable of universal computation there should be no loss of generality. One of the simplest approaches for enlarging the elementary 1*D* paradigm is to allow for varying neighborhoods which is simply incorporated by taking the kernel to be also time-dependent.

Following the general recipe for spectral composites in [6], the dual inverted sequence takes the total form

$$\tilde{\vec{h}}_{t+1} = \hat{Z}_K \cdot (F \circ R)(\vec{h}_t) \qquad (13)$$

Given the superpotential based reformulations of section 2, adopting a continuous embedding approach the generic $R$ mapping can now be replaced as for instance in the case of (7) with

$$\tilde{h}_{t+1} = \hat{Z}_K \cdot (F \circ c)(h_t) - \hat{Z}_K \cdot (F \circ c)(h_t - \delta x) \qquad (14)$$

Assuming a sufficiently smooth interpolated form for $c$ with an infinite periodic extension and following the method in [6], the two terms in (14) can be rewritten with the aid of an additional unitary kernel in the form

$$(F \circ c)(h_t)(k) = \int dk' \tilde{c}(k') U(k', k) \qquad (15)$$

The new kernel is then given by another FFT over $x$ parametrized by $k'$ as

$$U(k', k) = F_{k \leftarrow x}(\exp(2\pi i k' h(x))) \qquad (16)$$

The shift in the second term of (14) can still be extracted as a multiplier in which case we arrive at

$$\tilde{h}_{t+1}(k) = \hat{Z}_K \cdot \int dk' \tilde{c}(k') \left(1 - e^{2\pi i k' \delta x}\right) F_{k \leftarrow x}(\exp(2\pi i k' h_t(x))) \qquad (17)$$

Exchanging integrals allows rewriting (17) as a total Fourier integral using a new simplified kernel

$$W(h) = \int dk' V(k') e^{2\pi i k' h}$$
$$V(k') = \left(1 - e^{2\pi i k' \delta x}\right) \tilde{c}(k') \qquad (19)$$

The total operation in the dual picture of the activation field then becomes

$$\tilde{h}_{t+1} = \hat{Z}_K \cdot \left(F \circ W\left(F^{-1}(\tilde{h}_t)\right)\right) \qquad (20)$$

Noticing that the integral in (19) is only a parametric form of an inverse Fourier integral turned into a functional allows to promote the total quantity in a form of a spectral representation of the superpotential as a functional modulated by the input *h* in the abstract form

$$W(z) = \left(F^{-1}\left(1 - e^{2\pi i k' \delta x}\right)F\right)c(z) \qquad (21)$$

It should be mentioned that the use of the input as a modulation term for (21) reflects the original map exponentiation in the inverted domain as noticed in the previous section. Thus, a power tower of mappings finds its analog in successive functional compositions as the dual of a general recursion. Notably, the appearance of two shifted versions for the superpotential terms reminds of the Moiré technique [11], [12] of which the use as an abstract optical and natural computing element has drawn little attention as yet [13] [14].

In the previous only the simplest type of superpotential was examined. The same procedure could be followed for the interferometric version of (9) and (10) but with much more elaborate results since even approximating the hard sigmoid nonlinearity with *dk'/k'* would leave the inner unitary kernel to be expanded in a bi-infinite series of Bessel functions as $J_n(k')exp(2\pi i c(h))$. Integrals of $J_n(k')/k'$ are naturally splet between odd and even indices the first rapidly converging towards sigmoidal functions. The resulting spread over an infinite bandwidth is well known in studies of phase and frequency modulations [15], [16] and its influence in the dual picture of the activation field dynamics will be expanded upon in future work. Last but not least, a large body of previous theoretical work on nonlinear filters originating in old cybernetics has been omitted here including the so called, Wiener-Volterra analysis with multi-convolution filters [17], [18]. One of the disadvantages includes the exponential increase in the number of kernels required for a complete approximation of the nonlinearities involved.

**Discussion and conclusions**

A number of analytical reformulations of CA were examined in the effort to find efficient implementations on continuous, analog substrates. As much as in the original work in [5] it should be stressed that given appropriate encoding schemes, several cases of discrete dynamics could be emeddable in continuous media. When these correspond to universal computational models in the sense of Turing, they represent classes of possible alternative architectures for natural and optical computing. Moreover, the blend between continuous and discrete structures offers additional possibilities for the reexamination of natural processes as information carriers and symbolic manipulators. In future work, the possibility of programmable continuous media will be further explored using examples from the functional foundations of computer science.

**References**

[1] S. Wolfram, *"Cellular Automata as Models of Complexity"*, Nature **311**,5985 (1984) 419-424.

[2] P. Povalej *et al.*, *"Machine Learning with Cellular Automata"*, Int. Symp. Intel. Data Anal. IDA 2005: Advances in Intelligent Data Analysis VI, 305-315,

[3] O. Yilmaz, *"Machine Learning Using Cellular Automata Based Feature Expansion and Reservoir Computing."*, J. Cell. Aut., V0, (2015) 1- 38.

[4] O. Yilmaz, *"Symbolic Computation using Cellular Automata-based Hyperdimensional Computing"*, Neural Comp., **27**(12) (2015) 2661-2692.

[5] T. E .Raptis, *"Spectral Representations and Global Maps of Cellular Automata Dynamics"*, Chaos, Sol. Frac., 91, (2016)

[6] Bergner *et al.*, *"A Spectral Analysis of Function Composition and its Implications for Sampling in Direct Volume Visualization"*, IEEE Trans. Vis. Comp. Graph., **12**(5) (2006)

[7] V. Garcia-Morales, *"Universal map for cellular automata"*, Phys. Lett. A, **376** (2012) 2645-2657.

[8] V. Garcia-Morales, *"Symmetry analysis of cellular automata"*, Phys. Lett. A, **377** (2013) 276-285.

[9] V. Garcia-Morales, *"Origin of complexity and conditional Predictability in cellular automata"*, Phys. Rev. E **88**, (2013) 042814.

[10] J. Müller-Quade *et al.*, "*Algorithmic Design of Diffractive Optical Systems for Information Processing*", Physica D, **120**, (1998) 196-205.

[11] Oster, G., et al., "*Theoretical interpretation of Moire patterns*," J. Opt. Soc. Am. **54**(169) (1964)

[12] s. Yokozeki, "*Theoretical interpretation of Moire patterns*," Opt. Comm., **11**(4) (1974) 378-381.

[13] A. Asundi, K. H. Yung, "*Phase-shifting and logical moiré*", J. Opt. Soc. Am. **8**(10) (1991) 1591-1600.

[14] F. Mohammadi *et al.*, "*3D Machine Vision Employing Optical Phase Shift Moiré*", (2011) Innovative Computing Technology, Comm. Comp. Inf. Soc. Book Series, Springer.

[15] G. Lewis, (1997) "*Communications Technology Handbook*", 2nd Ed., Focal Press, Butterworth-Heinemann Linacre House, Oxford.

[16] A. Lozano-Nieto, (2011) *"Frequency and Phase Modulation"*, Hossein Bidgoli, Ed. In Chief, Handbook of Computer Networks: Key Concepts, Data Transmission, and Digital and Optical Networks, V. 1, Wiley.

[17] D. J. Kruzienski, K. Jenkins, *"Series Cascade Nonlinear Adaptive Filetrs",* Conf.: Circuits and Systems, 2002. MWSCAS-2002

[18] M. O. Franz, B. Schölkopf, "*A Unifying View of Wiener and Volterra Theory and Polynomial Kernel Regression*", Neural Comp. **18**(12) (2006) 3097 – 3118.

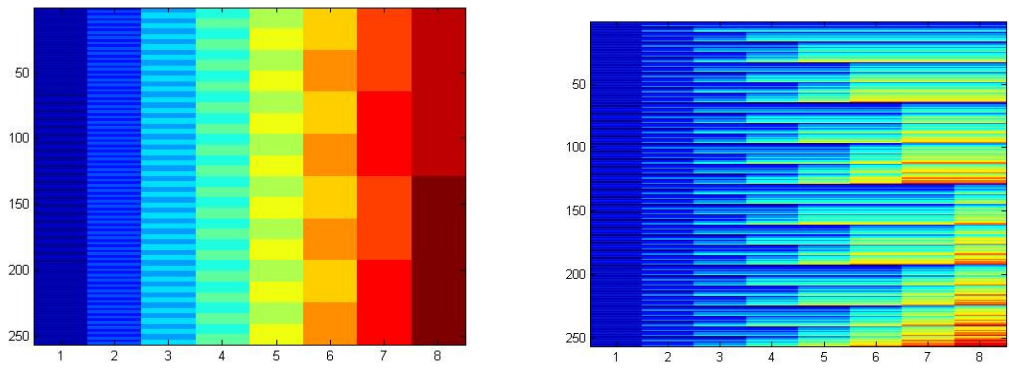

**Fig. 1**, The two alternative "integral" representations for all 256 elementary rules of (a) the ( *mod* 2) integral sequences and, (b) the cumulants over the original bit sequences.